\documentclass[a4paper]{VisionStyle}
\usepackage{graphicx}

\def\gin{{\it Ginga\/}}

\def\asca{{\it ASCA\/}}

\def\xmm{{\it XMM-Newton\/}}

\def\cha{{\it Chandra\/}}
\def\bep{{\it BeppoSAX\/}}
\def\int{{\it INTEGRAL\/}}

\def\H0{{\rm ~km~s^{-1}~Mpc^{-1}}}

\def\etal{et al.~\/}
\def\eg{{\it e.g.~\/}}

\def\la{\mathrel{\hbox{\rlap{\hbox{\lower4pt\hbox{$\sim$}}}{\raise2pt\hbox{$<$}}
}}}
\def\ga{\mathrel{\hbox{\rlap{\hbox{\lower4pt\hbox{$\sim$}}}{\raise2pt\hbox{$>$}}
}}}

\def\d25{D$_{\rm 25}$}

\def\.25{0.25 keV\thinspace}

\begin{document}
\title{Iron Features in the \xmm ~spectrum of \object{NGC 4151}}

\author{N.J.\,Schurch\inst{1} \and R.S.\,Warwick\inst{1} \and  R.E.\,Griffiths\inst{2} \and A.F.\,Ptak\inst{2}}

\institute{
Department of Physics and Astronomy, University of Leicester, University Road, Leicester, LE1 7RH
\and
Department of Physics, Carnegie Mellon University, 5000 Forbes Avenue, Pittsburgh, PA 15213 }

\maketitle 

\begin{abstract}

We present  a detailed  analysis of the  hard X-ray ($>$2.5  keV) EPIC
spectra from the first observations of  \object{NGC 4151} made by \xmm. We fit the
spectra with a model consisting  of a power-law continuum modified by 
line-of-sight absorption (arising  in both  partially  photoionized   
and  neutral  gas) plus additional iron-K emission and absorption  
features. This  model  provides  an excellent overall fit  to the 
EPIC ~spectra.   The iron K$\alpha$ line is well modelled  as a narrow 
Gaussian  component. In contrast to several earlier studies based on 
data from ASCA, a relativistically  broadened iron  K$\alpha$ emission 
feature is  not required by  the \xmm ~data. The upper  limit on  the flux
contained in  any additional  broad line is  $\sim$8\% of that  in the
narrow line. The measured intrinsic line width ($\sigma =32\pm7$  eV)
may be ascribed to (i) the 
doublet nature of the  iron K$\alpha$ line and (ii) emission from  low 
ionization states  of iron, ranging from neutral up  to ~FeXVII.  The 
additional iron absorption edge arises in cool material and implies a factor $\sim2$ overabundance of iron in this component.

\end{abstract}

\vspace{-4mm}

\section{Introduction}
 
The Seyfert 1 galaxy \object{NGC 4151}  is one of the brightest Active Galactic
Nuclei  (hereafter AGN)  accessible in  the X-ray  band, and  has been
extensively studied  by all  major X-ray missions.  This observational
focus has  revealed the  X-ray spectrum  of \object{NGC 4151}  to be  a complex
mixture  of  emission  and  absorption components,  originating  at  a
variety  of  locations  from  the  innermost  parts  of  the  putative
accretion  disk,  out  to  the  extended  narrow-line  region  of  the
galaxy.  The  X-ray spectrum  emanating  from  the  active nucleus  is
dominated by an intrinsic X-ray to $\gamma$-ray continuum that appears
to  be produced  by the  thermal Comptonization  of soft  seed photons
(e.g. \cite{Nschurch-C2:haa91}; \cite{Nschurch-C2:zdz94}, \cite{Nschurch-C2:zdz96}, \cite{Nschurch-C2:zdz00};
\cite{Nschurch-C2:pet00}),  plus additional contributions from reprocessing
in the form of Compton-reflection and features associated with neutral
iron (\eg iron-K fluorescence and a neutral iron edge). Below $\sim$ 5
keV the hard continuum is strongly cut-off by photoelectric absorption
in a  substantial ($N_H  \sim 10^{23} \rm~cm^{-2}$)  line-of-sight gas
column density (e.g. \cite{Nschurch-C2:hol80}; \cite{Nschurch-C2:yaq93}; \cite{Nschurch-C2:Wea94a}, \cite{Nschurch-C2:Wea94b}).

Some earlier studies based on \asca~ observations have concluded that
the line profile is complex, and is composed of with an intrinsically narrow component (\cite{Nschurch-C2:Wea94b}; \cite{Nschurch-C2:zdz94}, etc) plus a relativistically
broadened line feature  ({\it e.g.} \cite{Nschurch-C2:yaq93}; \cite{Nschurch-C2:Wan01}). Furthermore the  broad line profile was reported  to be variable
on  timescales of 10$^{4}$s,  corresponding to  an emitting  region of
$<$0.02  AU, suggestive  of  an origin close to the supermassive black hole
presumably in  the inner  regions of  a
putative accretion disk.

In  this paper we focus on the iron-K  features present in the 
very high signal-to-noise EPIC spectra of  \object{NGC 4151}. Our analysis is based on the
 spectral  ``template'' model, developed by \cite*{Nschurch-C2:Sch02} to interpret earlier \asca ~and \bep observations.

\section{The \xmm ~observations}

\object{NGC 4151} was observed with  \xmm ~three times during the 21$^{st}$ and
22$^{nd}$  of December,  2000  (orbit 190).  NGC  4151 was  positioned
on-axis  in both  the  EPIC MOS  and  PN cameras  (\cite{Nschurch-C2:Tur01};
\cite{Nschurch-C2:str01}) with the  medium filter in place. Of the three
EPIC observations, one ($\sim$33 ks) was performed with the MOS and PN
in small  window mode, whilst the remaining  two observations ($\sim$63
and  $\sim$23  ks)  were  operated   in  full  window  mode  for  both
instruments.  The  recorded  events  were  screened  with  the  latest
available  release of  the \xmm  ~Science Analysis  Software  (SAS) to
remove known hot  pixels and other data flagged as  bad. The data were
processed  using the  latest CCD  gain  values and  only X-ray  events
corresponding to patterns 0--12 in the  MOS cameras and 0--4 in the PN
camera were  accepted. Investigation of the full  field count-rate for
all  three observations  revealed a  single flaring  event  during the
first  observation. This  flaring event  was screened  from  the data,
resulting in effective total exposure times of 110 ks and 91 ks respectively
for the  MOS and PN instruments.  An investigation into  the impact of
pile-up on the observation showed that the effect was negligible in all the
observations, largely due  to the relatively faint nature  of \object{NGC 4151}
during this period.

A  source  and  background  lightcurve  were  extracted  from  each
individual,  screened  observation. These lightcurves revealed  that NGC  
4151 remained remarkably  constant (to
within  10\%) for  the duration  of the  observations, allowing  us to
analyse  the  \xmm  ~data   without  being  overly concerned  with  the
considerable  spectral variability  that  characterized the previous
 \asca  ``long-look'' observation (\cite{Nschurch-C2:Sch02}). Source  and background  
spectra were
extracted from the  same regions as the lightcurves.  The spectra from
the  two full  window mode  observations  were co-added  to produce  a
single  source spectrum  and  a single  background  spectrum for  each
instrument.  The spectra were  binned to  a minimum  of 20  counts per
spectral channel, in order to apply $\chi^2$ minimisation techniques.
Here we consider only the 2.5-12 keV spectrum.

\section{The spectral analysis}

\subsection {The initial spectral template model}

We  adopt the  spectral template  model  described by, \cite*{Nschurch-C2:Sch02} 
which includes the following emission components:

\begin{enumerate}
\item A  power-law continuum with  a normalization, $A_1$,  and photon
index, $\Gamma$, exhibiting a high-energy break at 100 keV;
\item A  neutral Compton-reflection  component (modelled by  PEXRAV in
XSPEC, \cite{Nschurch-C2:mag95}) with only  the reflection scaling
factor,  $R$, as  a free  parameter.  The parameters  relating to  the
incident  continuum   were  tied  to  those  of   the  hard  power-law
component.  In  addition  cos $i$  was  fixed  at  0.5 and  the  metal
abundance in the reflector was fixed at the solar value;
\item An iron K$\alpha$ emission line of intensity $I_{\rm K\alpha}$ at an
energy $E_{\rm K\alpha}$ with an intrinsic line width $\sigma_{\rm K\alpha}$.
\end{enumerate}

The  complex absorber is represented as the product of two absorption
components, namely a  warm column density $N_{\rm H,warm}$ and  a cold gas
column  $N_{\rm H,cold}$.  For  further  details  of  the  photoionization
modelling  see \cite*{Nschurch-C2:Sch02} and  \cite*{Nschurch-C2:gri98}. 
The adopted spectral  model includes absorption arising in the
line-of-sight column  density through our  own Galaxy, applied  to all
three emission components ($N_{\rm H,Gal}$= 2$\times10^{20}$ cm$^{-2}$).

\subsection{The spectral fitting}

\begin{figure}
\centering
\begin{minipage}{85 mm} 
\centering         \hbox{        \includegraphics[width=5.5        cm,
angle=270]{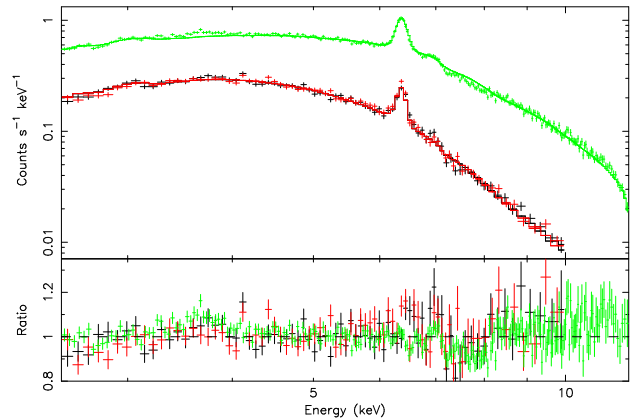}}
\caption{\scriptsize  The  \xmm ~EPIC  spectra  of  NGC  4151 and  the
spectral template model. The MOS data, shown here in black (MOS 1) and
red (MOS 2), are from the  small window mode observation. The PN data,
shown  here  in   green,  is  from  the  Full   Window  mode  co-added
observation.}
\label{nSchurch-C2_fig:spect1}
\end{minipage}
\end{figure}

\begin{figure}
\centering
\begin{minipage}{85 mm} 
\centering \hbox{ \includegraphics[width=5.5 cm, angle=270]{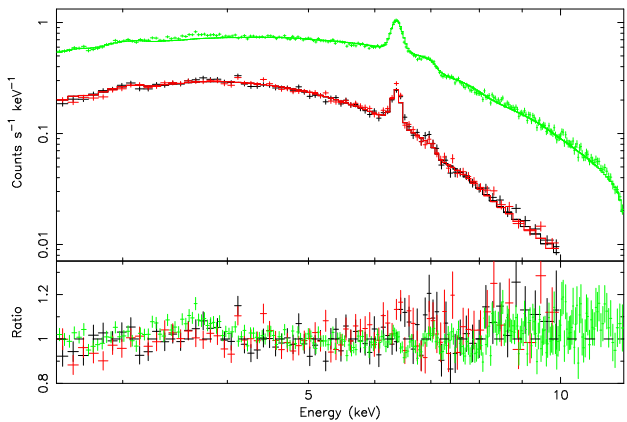}}
\caption{\scriptsize The \xmm  ~EPIC spectra of \object{NGC 4151}  fit with the
modified spectral template model. The EPIC spectra, MOS 1 (black), MOS
2 (red) and PN (green), are  fit with the spectral template model plus
an additional edge feature from neutral iron at 7.1 keV.}
\label{nSchurch-C2_fig:spect2}
\end{minipage}
\end{figure}

The template spectral  model was fit simultaneously to  both the small
window  mode and  the co-added,  large window  mode EPIC  spectra. The
initial model allowed the continuum and reflection normalizations, the
line  properties (centroid  energy, width  and normalization)  and the
ionization  parameter  of  the   warm  absorbing  column  to  be  free
parameters. The values for all the remaining model parameters were set 
at those used in \cite*{Nschurch-C2:Sch02}.
The  spectra, along with the spectral  template model fit,
is   shown  Figure   \ref{nSchurch-C2_fig:spect1}  and   the  best-fit
parameters  are given  in  Table \ref{nSchurch-C2_tab:fitnos1},  Model
1. The quoted errors (here as elsewhere in this paper) are at the 90\%
confidence level  as defined by  a $\Delta$$\chi$$^{2}$=2.71 criterion
({\it i.e.},  assuming one interesting  parameter). This model  gave an
acceptable  fit  to the  spectrum  ($\chi^{2}$=5465  for 4973  d.o.f),
although the  data/model residuals suggest an over-prediction of  the flux
between 7 and 8  keV. A closer inspection of the shape  of the data in
this region reveals a sharp drop at $\sim$7 keV, which we interpret as
an absorption edge from neutral  iron. Including an additional edge in
our  model,  with  the  edge  energy  and the  optical  depth  a  free
parameters, improved the  fit dramatically ($\Delta\chi^{2}$=173 for 2
d.o.f;  Figure \ref{nSchurch-C2_fig:spect2}). The  best-fit parameters
for  the model  incorporating an  additional edge  are given  in Table
\ref{nSchurch-C2_tab:fitnos1}, Model 2.

\begin{table}
\centering
\begin{minipage}{85 mm} 
\centering
\caption{Best-fit spectral model parameters}
\begin{tabular}{lcc}
Model & Model 1 & Model 2 \\
Parameter & & \\ \hline 
R & 0.58$^{+0.02}_{\rm -0.06}$ & 0.62$^{+0.03}_{\rm -0.06}$ \\
log($\xi$) & 2.631$^{+0.007}_{\rm -0.004}$ & 2.627$^{+0.003}_{\rm -0.007}$ \\ 
$A_{\rm pl}$$^{a}$ & 1.27$^{+0.03}_{\rm -0.01}$ & 1.28$^{+0.03}_{\rm -0.02}$ \\
$E_{\rm K\alpha}$$^{b}$ & $6.373^{+0.03}_{\rm -0.02}$ & 6.393$^{+0.003}_{\rm -0.002}$ \\ 
$\sigma_{\rm K\alpha}$$^{b}$ & 0.032$^{+0.07}_{\rm -0.07}$ & 0.032$^{+0.005}_{\rm -0.007}$ \\
$I_{\rm K\alpha}$$^{c}$ & $1.31^{+0.3}_{\rm -0.3}$ & 1.26$^{+0.03}_{\rm -0.03}$ \\
$E_{\rm K, Edge}$$^{b}$ & -  & 7.13$^{+0.04}_{\rm -0.03}$\\
$\tau_{\rm K, Edge}$ & -  & 0.11$^{+0.012}_{\rm -0.015}$\\
 & &  \\
$\chi^{2}$ & 5471.3 & 5297.7 \\
d-o-f & 4971 & 4969 \\ \hline
\multicolumn{2}{l}{\scriptsize $^{a}$ $10^{-2}\rm photon~keV^{-1}~cm^{-2}~s^{-1}$} & \multicolumn{1}{l}{\scriptsize $^{b}$ keV} \\
\multicolumn{2}{l}{\scriptsize $^{c}$ $10^{-4} \rm~photon~cm^{-2}~s^{-1}$} &  \\
\end{tabular}
\label{nSchurch-C2_tab:fitnos1}
\vspace{-4mm}
\end{minipage}
\end{table}

Including a  further contribution to  the iron K$\alpha$  line profile
from a broad ($\sigma$=0.2 keV) component does not improve the model fit, and sets an   
upper   limit  on   the   flux   in   the  broad   component   of
1.0$\times10^{-5}$ photons cm$^{-2}$  s$^{-1}$. Replacing the Gaussian
line in the  best-fit model with the broad  emission from an accretion
disk ({\bf disk-line} model in  XSPEC, \cite{Nschurch-C2:fab89}) results in a
similar fit to that for a narrow line ($\chi^{2}$=5309 for 4970 d.o.f).
However, in this case the derived inner radius  of the  accretion disk,  
r$_{\rm i}$, $\sim$1000r$_{\rm g}$ gives a disk-line profile that closely 
resembles the observed narrow line.

\section{Discussion}

The hard X-ray  spectrum of \object{NGC 4151} is well  modelled by the spectral
template  model   developed  in \cite*{Nschurch-C2:Sch02},  with
surprisingly few modifications.  In the \xmm~observations NGC  4151 
was in a relatively faint state as judged against previous extensive 
monitoring of the source by missions such as {\it EXOSAT} and {\it Ginga}. 
The  continuum is $\sim$60\% weaker, than the faintest level measured during earlier 
\bep ~and \asca ~observations, whereas the  level  of  the Compton  reflection
component  is $\sim$60\%  stronger.
The  ionization parameter of the warm absorber falls within  the  range of
ionization  states recorded  during the  \asca  ~long-look observation
($\xi$=2.48--2.65).

\subsection{The nature of the iron line}

Following  the analysis  described in  \cite*{Nschurch-C2:Sch02}, we
model  the iron  K$\alpha$ line  in the  \xmm ~spectra  with  a simple
Gaussian model  at an energy corresponding to  fluorescence of cold,
neutral iron. The  data are well modelled by this  line profile and we
find no evidence for the presence of a relativistically broadened iron
K$\alpha$ line. This is  in contrast with  some previous
claims (\cite{Nschurch-C2:yaq95}; \cite{Nschurch-C2:Wan01}) based  on the analysis of
\asca ~observations.  However, we note  that further analysis  of these
earlier data by \cite*{Nschurch-C2:Sch02} {\it do not} require the presence of
a broad  line. Including a broad  line in our modelling  gives no
statistical improvement  of the fit and  results in an  upper limit on
the flux  in the broad  component of $\sim$8\%  of that in  the narrow
component.  Modelling   the  line  profile   with  a  relativistically
broadened  disk-line profile with parameters values as quoted in 
\cite*{Nschurch-C2:yaq95}  results in a poor fit to the data
($\chi^{2}$=8800  for  4971  d.o.f).  Similarly,  modelling  the  line
profile with the  parameters quoted by \cite*{Nschurch-C2:Wan01} also results
in a  poor fit to the  data ($\chi^{2}$=9438 for 4971  d.o.f). In fact 
the broad line  profiles proposed by Wang \etal and Yaqoob \etal are 
both ruled out at $\gg$99.99\% significance.

The measured flux in the narrow iron $K\alpha$ line  is
1.26$^{+0.03}_{\rm -0.03}\times10^{-4}$    photons   cm$^{-2}$   s$^{-1}$,
significantly lower (at $\gg4\sigma$) than that measured in
the \asca, \bep ~and \gin ~observations where the line flux was more 
typically 2.2$^{+0.2}_{\rm -0.3}\times10^{-4}$ photons  cm$^{-2}$ s$^{-1}$; 
(\cite{Nschurch-C2:Sch02}, \cite{Nschurch-C2:yaq91})  the present measurment is 
also somewhat lower than the line flux
reported in more recent \cha ~observations (1.8$\times10^{-4}$ photons
cm$^{-2}$ s$^{-1}$; \cite{Nschurch-C2:Ogl00}), implying a possible decrease in the 
line flux over a period of $\sim$1 year. 

\xmm ~measures a  line width of
32$^{+7}_{\rm -7}$ eV, somewhat greater than the line width measured
by  \cha  ~($\sim$5  eV). Possible explanations for  the non-zero intrinsic width of the
Gaussian profile include:  (i) the signature of the  doublet nature of
the underlying iron K$\alpha$  line and (ii) emission from low  ionization 
states of
iron, ranging up  to ~FeXVII, present in the  warm absorber. However, modelling
the line in \xmm ~with two intrinsically narrow components to represent
the iron K$\alpha$  doublet yields a marginally  worse fit to the
data ($\chi^{2}$=5320  for 4970  d.o.f). The lines  were fixed  at the
correct   rest  energies   of  the   K$\alpha_{\rm 1}$  (6.392   keV)  and
K$\alpha_{\rm 2}$ (6.405 keV)  lines and the branching ratio  was fixed at
the theoretical  value of 1:2  respectively. We note that \cite*{Nschurch-C2:Ogl00} find no
evidence for Doppler broadening from their spectra.

\subsection{The iron edge - a clue to iron abundance.}

The  spectral model developed  in \cite*{Nschurch-C2:Sch02} includes
iron edge features from  both the neutral and warm absorbers and from 
the Compton reflection of neutral gas, assuming in each case
solar metal abundances. These  solar abundance edges  are insufficient to  
model the 7--8 keV region  of the \xmm ~spectrum, prompting the inclusion of an
additional edge. The derived edge energy is consistant with absorption by neutral iron. 
This feature can be interpreted either  as a
result of abundance gradients or as  a result of a general overabundance of
iron in the absorbing material.  Previous analyses of \object{NGC 4151}, albeit
using a  partial covering  model, have also  reported evidence  for an
overabundance  of  iron in  NGC  4151.  \cite*{Nschurch-C2:yaq93}  quote  a
canonical  value of  A$_{\rm Fe}\sim$2.5  times solar, based on the average 
of many measurements. Interpreting  the additional  edge required by  the \xmm
~data as an overabundance of iron in the neutral absorber (\cite{Nschurch-C2:mor83}), the optical 
depth of the edge leads
to a value for the iron abundance, A$_{\rm Fe}$, of 3.6$^{+2.15}_{\rm -1.8}$
times  solar. Alternativly, allowing the  iron abundance  of the  Compton reflection
component to  replace the additional edge  in the model  results in an
more modest iron  overabundance (A$_{\rm Fe}\sim$2.3), consistent with the
results of \cite*{Nschurch-C2:yaq93}, however in this case the fit becomes somewhat worse
above 8 keV. If we allow an overabundance in both the neutral absorption and in the 
neutral reflection component, a value of A$_{\rm Fe}\sim$2 is required.

\section{The Conclusions}

Remarkably, the  hard  X-ray (2.5--10  keV)  EPIC  spectra  from the  
recent  \xmm ~observation of \object{NGC 4151} are extremely well modelled by 
the model developed in \cite*{Nschurch-C2:Sch02} with only minor 
modifications. The success of this simple model in modelling both 
previous and current data supports that the spectral variability 
exhibited by \object{NGC 4151} is the product of rather subtle changes, specifically in the 
ionisation state of the majority of the line of sight warm absorber.
The parameters of the warm absorber fit to the \xmm ~data are consistent
with the  results from the  spectral analysis of  the past seven
years of \asca ~and \bep  ~observations. 

The chief modification to the template spectral model (\cite{Nschurch-C2:Sch02}) 
is the addition  of  an  edge  feature  from neutral  iron. Interpreting the 
edge as the result of an overabundance of iron in the cold absorber and cold 
reflector, then a value of A$_{\rm Fe}\sim$2 is required.

This analysis finds no requirement for a relativistically broadened
iron  K$\alpha$  line feature in the \xmm ~data. The upper limit on the possible  
flux present in any additional broad line is $\sim$8\% of that in the 
narrow line component. There is evidence that the line flux in \object{NGC 4151} has 
decreased recently, possibly due to the fact that the source has been fainter 
during the last few years than was previously observed. The low luminosity 
(L$_{\rm X,2-10}\sim$2$\times10^{42}$ erg s$^{-1}$) of the current observation is 
consistent with this view.

The intrinsic non-zero width of  the best-fit Gaussian line profile is
suggestive of a further level of complexity to the iron K$\alpha$ line
feature, with possible explanations  for the non-zero intrinsic width 
including: (i)  the signature of  the doublet nature  of the
iron K$\alpha$ line, (ii) emission from low ionization states of iron,
ranging up to ~FeXVII, present in the warm absorber.

\end{document}